\documentclass[
    journal,
    twocolumn,
    ]{IEEEtran}
\usepackage{graphicx}
\usepackage{amsmath}
\usepackage{amssymb}
\usepackage{epsfig}
\usepackage{epstopdf}
\usepackage{amsthm}
\usepackage{afterpage}
\usepackage{amsmath,amssymb,amsfonts}
\usepackage{verbatim}
\usepackage{psfrag}
\usepackage{color}
\usepackage{cite}
\usepackage{setspace}
\usepackage{adjustbox}
\usepackage{epsfig}
\usepackage{epstopdf}
\usepackage{footmisc}
\usepackage{fmtcount}
\usepackage{stackrel}
\usepackage{float}
\usepackage{breqn}
\usepackage{hyperref}
\usepackage{enumerate}
\usepackage{graphicx}
\usepackage{subcaption}
\usepackage{graphicx}

\DeclareMathOperator*{\argmin}{arg\,min}
\usepackage{tabularx,booktabs}

\begin{document}

\title{Bayesian Multidimensional Scaling for Location Awareness in Hybrid-Internet of Underwater Things}
\author{Ruhul Amin Khalil, \IEEEmembership{Member, IEEE;} Nasir Saeed, \IEEEmembership{Senior Member, IEEE;} Mohammad Inayatullah Babar, Tariqullah Jan and Sadia Din, \IEEEmembership{Member, IEEE} 
\thanks{Ruhul Amin Khalil, Mohammad Inayatullah Babar and Tariqullah Jan are with the Department of Electrical Engineering, Faculty of Electrical and Computer Engineering, University of Engineering and Technology, Peshawar 25120, Pakistan e-mail: [1. ruhulamin, 2. babar, 3. tariqullahjan]@uetpeshawar.edu.pk.}
\thanks{Nasir Saeed is with Department of Electrical Engineering, Northern Border University, Arar 73222, Saudi Arabia e-mail: mr.nasir.saeed@ieee.org.}
\thanks{Sadia Din is with the Department of Information and Communication Engineering, Yeungnam University, Gyeongbuk 38541, South Korea email: sadiadin@yu.ac.kr.}
\thanks{Manuscript received ~X, X~X ; revised ~X, X~X, accepted ~X, X~X.}}
\markboth{IEEE/CAA JOURNAL OF AUTOMATICA SINICA,~Vol.~X, No.~X, X~X}%
{Shell \MakeLowercase{\textit{et al.}}: Bare Demo of IEEEtran.cls
for Journals}

\maketitle

\begin{abstract}
Localization of sensor nodes in the Internet of Underwater Things (IoUT) is of considerable significance due to its various applications, such as navigation, data tagging, and detection of underwater objects. Therefore, in this paper, we propose a hybrid Bayesian multidimensional scaling (BMDS) based localization technique that can work on a fully hybrid IoUT network where the nodes can communicate using either optical, magnetic induction, and acoustic technologies. These communication technologies are already used for communication in the underwater environment; however, lacking localization solutions. Optical and magnetic induction communication achieves higher data rates for short communication.
On the contrary, acoustic waves provide a low data rate for long-range underwater communication. The proposed method collectively uses optical, magnetic induction, and acoustic communication-based ranging to estimate the underwater sensor nodes' final locations.  Moreover, we also analyze the proposed scheme by deriving the hybrid Cramer-Rao lower bound (H-CRLB). Simulation results provide a complete comparative analysis of the proposed method with the literature.
\end{abstract}

\section{Introduction}
\label{sec:introduction}
\IEEEPARstart{I}{nternet} of underwater things (IoUT) has attained much attention because of their many applications such as navigation, objects localization, detection of mines, and monitoring of environmental pollution \cite{khalil2020towards,ali2020recent,kao2017comprehensive,zhao2020privacy}. The growth of IoUT is in its early phase and faces innumerable challenges. For example, in terms of communications, the radio frequency (RF) waves do not provide satisfactory outcomes due to various factors such as scattering and absorption in an underwater environment. In contrast, acoustic waves are vastly utilized for underwater communication due to its low-absorption in water \cite{domingo2012overview,gussen2016survey}. The acoustic waves can travel for long distances up to tens of kilometers, but offers a low data rate and have a substantial propagation delay \cite{schirripa2020underwater}. Therefore, to provide better data rates and a little propagation delay, optical communication is recently used to develop optical-IoUT networks. Optical communication is utilized to correspond among the underwater things for shorter range and consists of high-quality light-emitting diodes (LEDs) or lasers. Optical communication provides higher data rates compared to its counterpart technologies \cite{r11,khalil2020effect}. However, propagation of optical light in the underwater environment is profoundly affected by the inherent properties of light, type of water, salinity, and turbulence \cite{al2018optical}. Nevertheless, magnetic induction (MI) is also utilized for underwater communication to provide high data rates. However, MI has a limited transmission range and is also affected by the conductive nature of the water\cite{khalil2020optimal,r7,wei2020dynamic}. 

Besides communication, localization in IoUT is of significant importance for tracking various underwater sensor nodes, data tagging, and detection of underwater targets \cite{khalil2020towards}. Moreover, the classical localization techniques for terrestrial IoT networks do not function well in marine environments due to the harsh nature and non-availability of the Global Positioning System (GPS) system. It is straightforward that GPS works well in terrestrial networks, but its performance degrades when used in an indoor \cite{yassin2016recent, SaeedMDSSurvey} and underwater environment \cite{r9}. Moreover, the underwater monitoring systems demand accurate localization techniques as the collected data is only useful if the nodes' location is estimated accurately \cite{saeed2018energy, saeed2020around}. Based on the communication technology used, underwater localization techniques can be divided into acoustic, optical, and MI-based systems. Various localization techniques for acoustic underwater sensor networks' been investigated in the past. These techniques consider different aspects of the system such as signal propagation model, network topology, environmental factors, localization accuracy, number of anchor nodes, the geometry of anchor nodes, and the sensor node's relative location to the anchors \cite{teymorian20093d,liu2010asymmetrical}. Most of the acoustic-based underwater localization systems use time difference of arrival (TDoA) ranging. However, the TDoA measurements for distance estimation in underwater acoustic communication channels are highly affected by multi-path. Similarly, the RSS-based distance estimation also suffers from multi-path propagation, making it hard to compute accurate distance estimation \cite{su2020review}. Nevertheless, the underwater acoustic channels show good transmission features at certain depths where RSS-based distance estimation can be a good option \cite{hosseini2011new}.
      
In case of underwater optical wireless communication, the optical light mainly suffers from attenuation, scattering, and absorption \cite{Nasir2018limited,akhoundi2017underwater}. Based on these impediments, the existing literature only presents the time of arrival (ToA) and RSS-based underwater localization schemes. For instance, \cite{akhoundi2017underwater} proposes an underwater optical positioning system using both  ToA and RSS-based ranging, where optical base stations (OBS) are used as anchor nodes. The sensor nodes receive the transmitted optical signal from various anchor nodes and locate themselves by utilizing a linear-least-square solution.  In contrast, a distance estimation technique based on RSS has been presented in \cite{arnon2009non} for a given underwater optical communication network. This technique strongly relies on various parameters that include properties of the optical channel, transmitted power, angle of divergence of the transmitter, trajectory angle, and field view of the receiver.  Some recent works also use Magnetic Induction (MI) for underwater communication and localization.  \cite{lin2017magnetic} proposes a novel MI-based localization technique that utilizes received magnetic field strength (RMFS) for measuring RSS. The proposed system considered proper elimination of multi-path fading and utilization of constant properties of the MI channel. 

Hence, many location awareness methods have been developed in the past to accurately estimate the position of the underwater nodes using different communication technologies. Likewise, there exist some other localization techniques that utilize the fusion of acoustic and optical communication technologies in finding the position of unknown nodes. Recent studies show that the hybrid approach can achieve better accuracy and have more network flexibility. Therefore, in this paper, a fully hybrid location awareness algorithm has been proposed, that considers three different technologies including MI, acoustic and optical communication. Such an approach where the system can utilizes any available technology, is also termed as a signals of opportunity (SOA) based localization. The main contributions of this paper are summarized as follows:
\begin{itemize}
\item A SOA based hybrid approach is introduced for localization of IoUT devices in the underwater environment, which takes into account acoustic, optical, and magnetic induction based ranging. 
\item BMDS-based dimensionality reduction technique, along with Procrustes analysis, is used to estimate the position of underwater IoUT devices. The proposed method uses the ranges from the received power using either optical, MI, or acoustic communication.
\item A  hybrid Cramer-Rao, lower bound (H-CRLB) is derived for the analysis of the proposed scheme. Numerical results show the effectiveness of the proposed method, where it achieves the H-CRLB.
\end{itemize}

The rest of the paper is characterized as follows. Section II presents a brief survey of the literature on underwater localization. In Section III, we present the BMDS based system modeling for localization in an IoUT and study its effectiveness for a hybrid optical-acoustic-magnetic induction approach. In Section IV, performance evaluation of the proposed system is carried out, followed by conclusions in Section V.

\section{Related Work and Challenges}\label{sec2}
\subsection{Literature background}
The localization algorithm's for wireless networks are divided into three categories. The first category involves algorithms that are either centralized or distributed \cite{r11}. In centralized algorithms, ranging measurements from all the nodes are shared with a central node to find their respective location. On the contrary, in distributive algorithms, each node is capable enough to locate itself with the help of available anchors. The second category includes range-based and range-free localization schemes. The range-based localization methods usually work with the measurements priorly available among the nodes to accurately locate unknown nodes \cite{r11}. In contrast, the range-free schemes only consider the proximity information and can work effectively without requiring the actual ranging measurements. Whilst the range-free schemes are simple and easy in a design perspective, but provide less accuracy in position estimation as compared to range-based techniques. Range-based location awareness techniques usually necessitate different ranging measurements such as angle of arrival (AoA), ToA, TDoA, and RSS \cite{r12}. The third category solely rely on the availability of anchor nodes, i.e., anchor-based and anchor-free schemes for localization. The anchor-based schemes require the availability of at least three anchor nodes in a 2D-space for a node to be able enough to localize itself. Oppositely, the anchor-free methods do not require anchor nodes for self-localization. It only utilizes the estimated local distances among the nodes to locate the unknown node effectively. A technique known as the linear least square solution is applied to further refine the estimated position of the node. \par 

In the range-free localization schemes, the most critical and well-known methods applied are usually based on the reduction of dimensionality. BMDS is considered as one of the fundamental dimensionality reduction techniques that are utilized extensively for localization \cite{r14,bakker2013bayesian,wang2020localization,lin2019bayesian}. Some other applications of BMDS include socioeconomics, political science, statistical economics, and behavioral sciences. Besides, BMDS is also used for the localization of nodes in IoT networks. According to \cite{r16}, the connectivity problem between two nodes is addressed by BMDS for the precise geometrical representation of the connecting nodes. Initially, BMDS is used to calculate the local maps of the sensor nodes and are then combined to obtain the global mapping. Manifold learning is another localization technique used for position estimation in wireless sensor networks \cite{r19}.  Applying BMDS for the localization of a hybrid IoUT system is not straightforward since multiple range measurements are available, which need to be integrated first before utilizing it for location awareness in the given underwater environment. \par

In this paper, we have utilized a centralized RSS-based ranging algorithm with the presence of anchors. First of all, centralized network localization schemes have better localization accuracy in harsh environments, such as underwater or underground. In centralized ranging, the end-user does not require to localize itself due to limited resources. Usually, its location is computed periodically by either the surface buoy or sink node. Secondly, we considered a range-based mechanism as it has better accuracy compared to range-free schemes \cite{pandey2016range}. Moreover, the logic behind considering the RSS-based approach is that the received power in communication systems is already available, and based on the channel, it can be converted into estimated distances. Since we are considering a hybrid model, time-based or angle-based ranging can lead to extra complexity and cost of the system \cite{SaeedMDSSurvey}.

Recently, BMDS-based localization for underwater applications has gained much importance due to its robustness and accuracy in position estimation. The marine environment is quite harsh, turbid, and hazy due to which robust optimization techniques need to be used to locate the position of IoUT devices accurately. Many variations of BMDS-based localization schemes are used in the literature, such as unconstrained optimization conjugate gradient method for multi-hop underwater networks based on optical medium. To the best of authors knowledge, none of the existing works tackle the problem of a fully hybrid IoUT network localization. Therefore, this paper provides an BMDS-based localization scheme for a hybrid Magnetic Induction-Acoustic-Optical (MIAO) underwater communication system that works on a SOA approach. The proposed scheme is novel in the context that it is used for the very first time in an underwater environment to localize IoUT devices. Furthermore, Procrustes analysis provided in \cite{r21} is applied to precisely estimate the final position of the nodes in the given hybrid system. Later on, an H-CRLB is also derived to define the lower threshold for position estimation of the proposed system. Simulation results are carried out in Matlab, to show that the proposed model is efficient and provides greater accuracy in position estimation using SOA approach. The next section provides a detailed layout of the network model and proposed technique for efficient localization in the IoUT.

\begin{figure*}[htp!]
\centering{\includegraphics[width=0.925\textwidth]{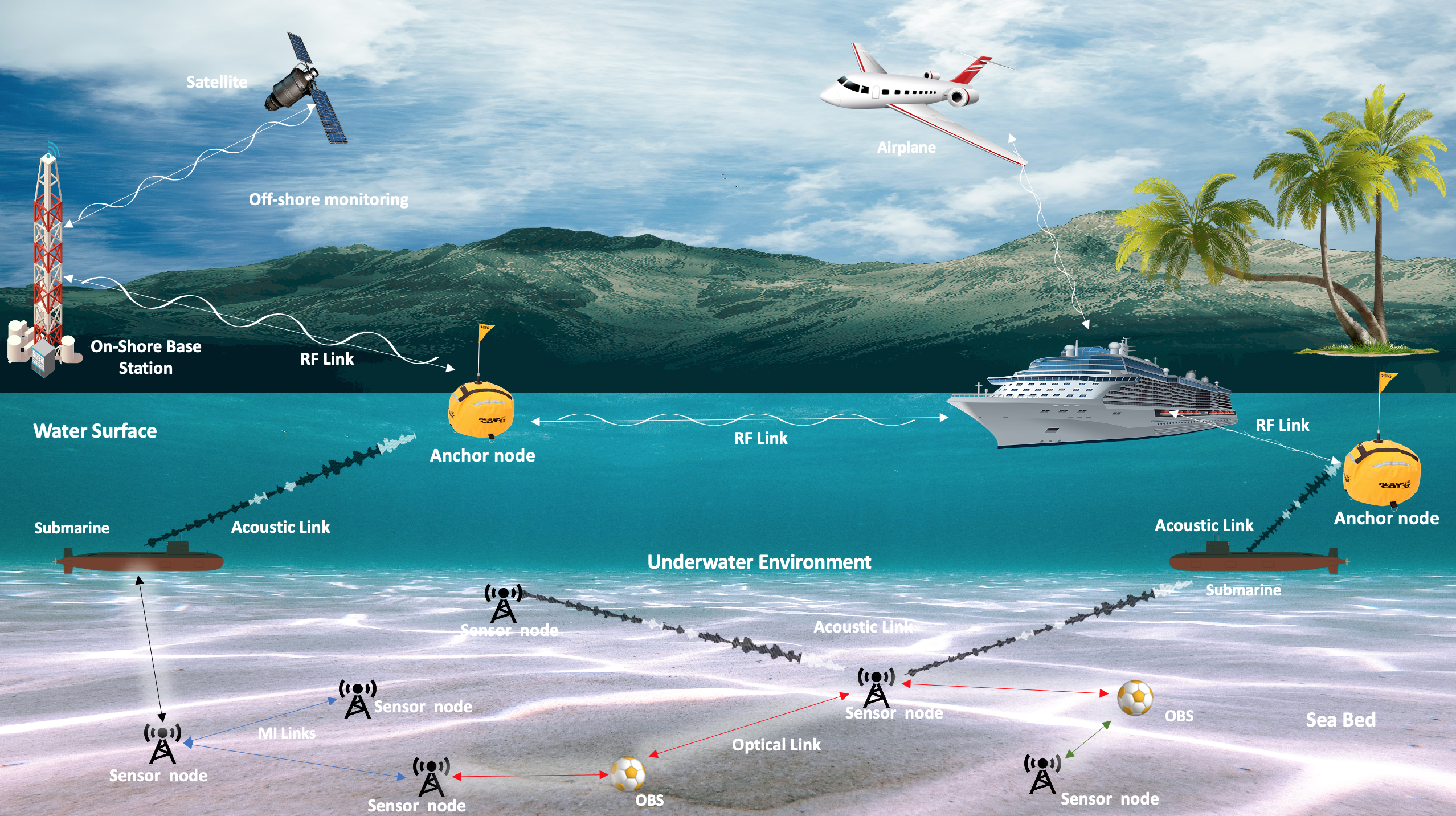}}
\caption{Proposed model for Hybrid-Internet of Underwater Things (H-IoUT)\label{proposed}}
\end{figure*}

\subsection{Challenges in Hybrid-IoUT}
   Although the hybrid approach brings new opportunities, it also poses many challenges that need to be considered when performing localization in the underwater environment. Some of these challenges include variations in seawater characteristics, misalignment for optical links, latency in acoustic connections, and coil orientation mismatch in magnetic links \cite{khalil2020towards}. The acoustic and optical links undergo severe performance degradation when impacted by seawater's refractive and composition properties. The refractive index changes occur due to variation in temperature, turbulence, and salinity, while the composition of water depends upon changes in the geographic location. Moreover, in acoustic links, increasing the transmission power can maintain a certain SNR level; however, several other factors degrade the SNR in optical links, such as attenuation-and-scattering, misalignment, divergence angle, and field of view. Analogously, a major issue with acoustic links is higher latency that can be minimized using multi-hop networking; however, it cannot be fully eliminated and must be considered when designing a hybrid underwater model.

   On the other hand, MI systems require proper orientation of the transceivers' coils in the harsh underwater environment. Although this can be achieved at the deployment stage, however, the transceivers certainly drift with ocean tides, and currents get rotated to undesired directions. Another critical factor for underwater MI communication is the eddy current loss due to the magnetic field. For instance, MI as a carrier in an aquatic environment can only achieve a range up to 30m at an operating frequency of 500 Hz \cite{domingo2012overview}. 
   
  Therefore, we propose a hybrid MIAO model where each communication technology supports the other, resulting in a better communication and localization performance. Nevertheless,  integration of these various technologies in the underwater environment is quite a daunting task. Integration of multiple communication systems requires a combination of electronics components, sub-systems, chip-level assembly with specific system functionality. The proposed model can foster interaction among numerous disciplines that will impact electronics, photonics, electromagnetism, and underwater communication systems. Consequently, there is a need to address the scientific issues and challenges associated with the underpinnings of systems integration in hybrid communication networks. The aim here is to achieve effective underwater communication and localization utilizing these different available underwater communication systems.

\section{Network Model and Ranging}\label{sec3}
This section first describes the network model, then introduces the novel concept of MIAO ranging, and finally discuss the use of BMDS for IoUT localization.

\subsection{Network Model}\label{subsec3.1}
Consider an IoUT network that is composed of $M$ anchor nodes and $N$ sensor nodes. These nodes are presumed to be lying and embedded on the sea bed, and few of them are suspending sensor nodes. Fig. \ref{proposed} depicts a generalized overview of the proposed network where the underwater objects with unknown location are communicating using various technologies.  Note that, the system is hybrid and can use any of the available optical, magnetic induction, and acoustic technologies for communication among them and with the surface buoy. Also, $M<N$, and the location of each anchor node is a well-known prior. According to the proposed methodology, it is assumed that each underwater object is able enough to communicate with each other object in the network through at least one path (connected network). A surface buoy is needed to collect and share the respective information shared by every node of the network.

The proposed algorithm comprises three major steps defined as follows;
\begin{enumerate}[(a)]
\item Each sensor node attempts to search for the overall neighbourhood by utilizing any of the communication technology and estimate the range to the adjacent nodes. 
\item Some sensor nodes are not lying in the communication range of each other and can utilize the available connectivity information and estimate the missing pairwise ranges.
\item The information from each node is communicated to the surface buoy through SOA approach. The initial ranging of every node is carried out on SOA approach that utilizes the characteristic of any available underwater links such as acoustic, optical, MI or hybrid. This approach enhances the estimation of distance among available nodes. The surface buoy utilizes the information provided and calculates the estimated distance matrix in a pairwise manner. It further applies the dimensionality reduction methodology based on subjective manifold interpretations to accurately locate each sensor node. 
\end{enumerate}
In the following, we describe the different types of ranging methods and show how can they be integrated into the SOA concept.

\subsection{Underwater Magnetic-Induction Ranging}\label{subsec3.2}
\begin{figure}[h!]
\begin{center}
\includegraphics[width=1\columnwidth]{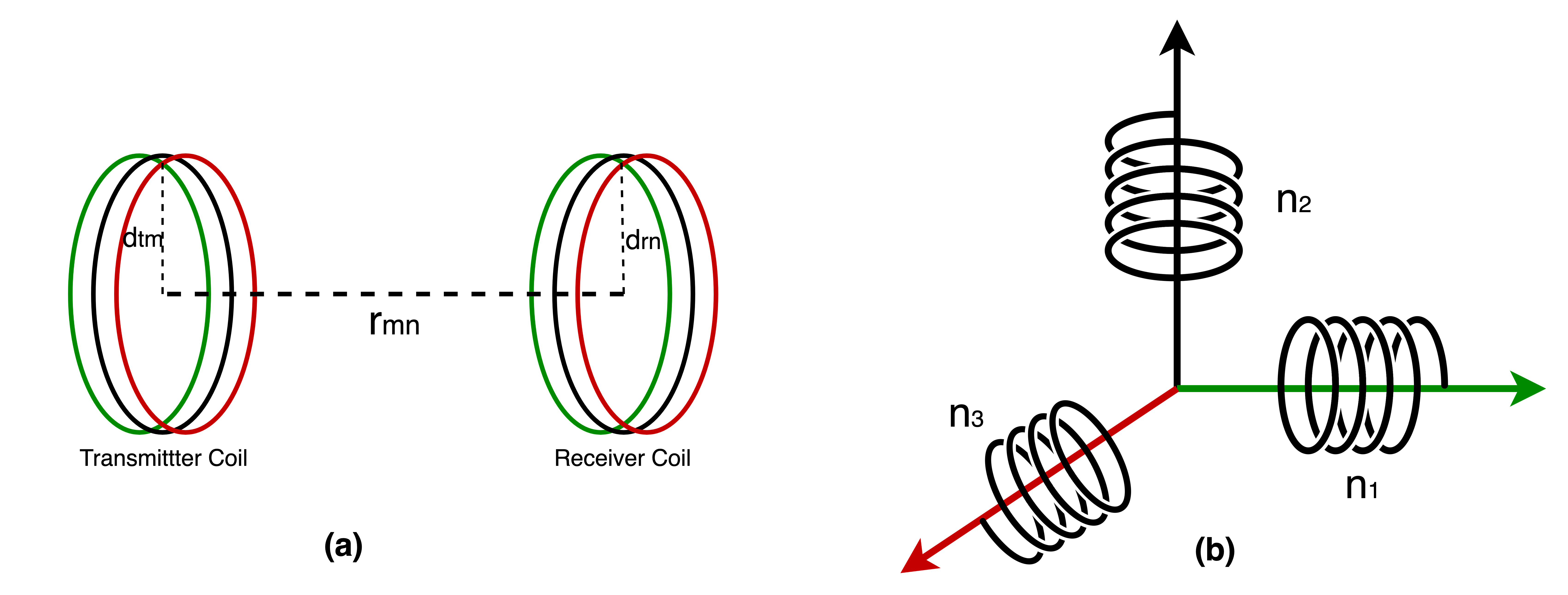}  
\caption{(a) MI link for underwater communication (b)Tri-directional coil antenna.\label{Fig2}}  
\end{center}  
\end{figure}

The exchange of information among various anchor and sensor nodes in a MI-based IoUT truly rely on the time-varying magnetic field. The time-varying magnetic field is basically generated by the modulated signal transmitted from transmitter coil antenna \cite{li2020underwater,muzzammil2020fundamentals}. It is also responsible for the intercommunication among the anchor and sensor nodes using magnetic induction (MI) phenomena as a medium in IoUT. As, the fabrication of the time-varying magnetic field is carried out by transmitter coil antenna through a sinusoidal modulated signal and induces current in the receiver coil antenna of the receiver. The induced current is responsible for demodulation of the signal to retrieve the embedded information. 
The realization of an MI-based transceiver is depicted in Fig. \ref{Fig2}. 
Assume that the current flowing through the transmitter coil antenna is represented by
\begin{equation}\label{eq1}
\begin{aligned}
I = I_{0}  \text{exp}^{-j\omega t} 
\end{aligned}
\end{equation}
where $I_0$ is termed as the direct current, the term $\omega$ in the superscript represents the angular frequency, where $t$ shows the instantaneous time. The direct current $I_0$ is responsible for inducing the current in the neighboring coil antenna by a phenomenon known as mutual induction. It should be noted that mutual induction phenomena do not result well if the transmitter and receiver coil antennas are not accurately coupled. As, the underwater environment is harsh; therefore the assumption of a tri-directional coil is used \cite{r22}. The tri-directional coil antenna can receive strong induced signals if there comes little variation in the coupling formation and thus making it omnidirectional. 

The coils in the tri-directional antenna are mutually orthogonal, and hence, do not interfere with the reception of each other. The signals received at any of these mutually orthogonal coils are combined at the receiver to demodulate the required information. This technique of using tri-directional antenna essentially reduces the misalignment factor that affect the overall communication \cite{r22}. According to \cite{r23}, the MI-based relationship between the power transmitted $P_{MI_{t_m}}$ and power received $P_{MI_{r_n}}$, for some high range of frequencies and large number of turns $Z_{t_m}$ in coil antenna of the transmitter is expressed by
\begin{equation}\label{eq2}
\begin{aligned}
P_{MI_{r_n}} = \frac{\omega\mu P_{MI_{t_m}} Z_{r_{n}} Z_{t_{m}} {d_{t_{m}}^{3} {d_{r_{n}}^{3} \sin^{2}{\vartheta_{mn}}}}}{16 D_{0} r_{MI_{mn}}^{2}},
\end{aligned}
\end{equation}
in \eqref{eq2}, the term $\mu$ provides the permeability of water, where $d_{t_{m}}$ is characterized as diameter of the transmitter coil antenna. Also, the term $Z_{r_{n}}$ shows the number of turns in the receiver coil, and $d_{r_{n}}$ is the respective diameter of the receiver coil antenna. $\vartheta_{mn}$ is the angle between the axis of transmitter and receiver coil antennas, $D_0$ is the impedance of unit length loop, and $r_{MI_{mn}} = {\vert\vert} \alpha_{m} - \alpha_{n} {\vert\vert}$ is the Euclidean distance between the transmitting and receiving coil antennas. It should be noted that path loss in \eqref{eq2} do not consider the phenomena of skin depth. According to \cite{sun2013optimal,r230}, the received power for  MI channel is defined as
\begin{equation}\label{eq4}
\begin{aligned}
10^{\frac{P_{{MI}_{r_n}}}{10}} = 10^{\frac{(P_{MI_{t_n}} - L_{MI})}{10}} + S,
\end{aligned}
\end{equation}
where $\ell_{MI}$ [dB] is the path loss and $S$ is the Gaussian distributed random variable with zero mean and standard deviation of $\phi$. In case of $N$ number of received magnetic field strength measurements, i.e., $P_{{MI}_{r_1}}, \cdots, P_{{MI}_{r_N}}$, which are identical and independent Gaussian variable with mean ${\Theta}_{MI}$ and variance $\phi^{2}$. Then, the likelihood function $\ell(\cdot)$ be can written as 
\begin{equation}\label{eq:LMI}
\begin{aligned}
\ell(\Theta_{MI} | P_{{MI}_{r_1}}, \cdots, P_{{MI}_{r_N}}) \hspace{1.5in} \\ 
= \prod_{n=1}^{N} \frac{\exp(-(P_{{MI}_{r_n}}-\hat{\Theta}_{MI})^{2}/2\phi^{2})}{\sqrt{2\pi\phi^{2}}}
\end{aligned},
\end{equation}
where $\hat{\Theta}_{MI} = \frac{1}{N}\sum_{n=1}^{N}P_{MI_{r_n}}$. By using the unbiased estimator $\hat{\Theta}_{MI}$, $r_{MI_{mn}}$ is estimated as \cite{r230}
\begin{equation}\label{eq3}
\begin{aligned}
\tilde{r}_{MI_{mn}} = f(P_{MI_{r_{n}}}) \hspace{1.9in} \\
= \argmin \Bigg\{ r_{MI_{mn}} \Bigg{|} \bigg(10^{\frac{(P_{MI_{t_m}} - \hat{\Theta}_{MI})}{10}} -1\bigg) \Bigg\}.
\end{aligned}
\end{equation}
Solving \eqref{eq3} yields
\begin{equation}\label{eqMI3}
\begin{aligned}
\tilde{r}_{MI_{mn}}= \frac{16 D_{0_{t}} D_{0_{r}} r_{MI_{mn}}^{3}} {{\omega^{2}{\mu^{2} Z_{t_{m}} Z_{r_{n}} {d_{t_{m}}^{3} {d_{r_{n}}^{3}} G^{2}(\sigma, \omega, r_{MI_{mn}})}}}}
\end{aligned}.
\end{equation}
the term $G^{2}(\sigma, \omega, r_{MI_{mn}})$ represents the loss occurred due to skin depth, where $\sigma$ gives the electrical conductivity of water. It is observed that the values of $\sigma$ changes with the types of water. For instance, the value of $\sigma$ = 0.01 S/m for clean water, but it goes to a higher value of $\sigma$ = 4 S/m for turbid sea water. These performance degrading factors of the MI technique in different types of underwater environment needs the coil configuration to be selected carefully in order to avoid the losses.

\subsection{Underwater Acoustic Ranging}\label{subsec3.3}
Acoustic ranging works well even for a larger distance in an underwater environment as compare to MI. The acoustic channel in underwater medium experiences two types of losses during  propagation: delay spreading and path loss attenuation \cite{r24}. The delay spread is caused by the combined effect of cylindrical and spherical losses, while the path loss attenuation occurs due to multiple effects such as absorption, scattering, leakage from ducts and diffraction \cite{r25}. Usually, the path loss function in an underwater environment between any two nodes $m$ and $n$ can be computed using the following equation
\begin{equation}\label{eq6}
\begin{aligned}
\upsilon_{t} = \upsilon_{c} + \upsilon_{s} + 10^{-3} \phi r_{A_{mn}} ,
\end{aligned}
\end{equation}
where  $\upsilon_{c}$ is the cylindrical loss, $\upsilon_{s}$ is the spherical loss,  $\phi$ is respective coefficient of absorption, and $r_{A_{mn}}$ is the corresponding euclidean distance between the two nodes for acoustic communication. According to Thorp model for absorption given in \cite{r26}, the factor $\phi$  only depends on the frequency $f$ of the transmitted signal given by
\begin{equation}\label{eq7}
\begin{aligned}
\phi = \frac{0.11f^{2}}{1+f^{2}} + \frac{44f^{2}}{4100+f^{2}}
\end{aligned}.
\end{equation}
The spherical spread loss works effectively in underwater environment and; therefore, ignoring the loss due to cylindrical spreading, we achieve
\begin{equation}\label{eq8}
\begin{aligned}
\upsilon_{t} = 20\log(r_{A_{mn}}) + 10^{-3} \left[\frac{0.11f^{2}}{1+f^{2}} + \frac{44f^{2}}{4100+f^{2}}\right]r_{A_{mn}}
\end{aligned}.
\end{equation}
using \eqref{eq8}, the distance is estimated from  $\tilde{r}_{A_{mn}}$  by utilizing only the real term of the Lambert function $W_0$  as follows \cite{r27}
\begin{equation}\label{eq9}
\begin{aligned}
\tilde{r}_{A_{mn}} = \left[ \frac{(2\times10^{4}) \hspace{0.015in} W_{0}(1.15\text{exp}^{-4} \hspace{0.015in}\phi \hspace{0.015in}\text{exp}^{0.11\upsilon_{t}})}{2.3 \hspace{0.015in} \phi} \right]
\end{aligned}
\end{equation}

\subsection{Underwater Optical Ranging}\label{subsec3.4}
Optical communication in underwater mostly suffers from a number of parameters such as scattering, absorption, angular attenuation, and widening \cite{r28}. It has been noticed that angular attenuation and widening of the optical signals during underwater communication is purely dependent on the wavelength of the transmitting signal. Conferring to \cite{r29}, the loss coefficient $l(\lambda)$ is based on the coefficient for scattering $s(\lambda)$ and absorption $a(\lambda)$, which is given by
\begin{equation}\label{eq10}
\begin{aligned}
l(\lambda) = s(\lambda) + a(\lambda)
\end{aligned}.
\end{equation}
also, the loss occurrence $L_{mn}$ during underwater propagation depends upon the euclidean distance $r_{O_{mn}}$ and loss coefficient $l(\lambda)$. Their mutual relationship is expressed by
\begin{equation}\label{eq11}
\begin{aligned}
L_{mn} = \text{exp}^{-l(\lambda)r_{O_{mn}}}
\end{aligned}.
\end{equation}

Here, we assume line-of-sight (LOS) underwater optical communication, where the node $m$ directly communicate with node $n$ using the optical light. According to \cite{r30}, the power received $P_{O_{r_n}}$ at node $n$ as a result of LOS communication with node $m$ is given by
\begin{equation}\label{eq12}
\begin{aligned}
P_{O_{r_n}} = P_{O_{t_m}}\eta_{m}\eta_{n}L_{mn} \left[ \frac{A_{n}\cos{\theta}}{2\pi r_{O_{mn}}^{2}(1-\cos\theta_{0})}\right]
\end{aligned},
\end{equation}
where $P_{O_{t_m}}$ represents the total power transmitted by node $m$, whereas  $\eta_m$ and $\eta_n$ are the optical efficiencies of node $m$ and $n$, respectively. The aperture area of the $n$ is given by $A_n$. Also, there exists various angles such as the angle of aperture between the trajectories of node $m$ and node $n$, that is represented by $\theta$, moreover the angle of divergence of the transmitted signal is given by $\theta_{0}$. In literature, many modulation techniques are used for optical wireless communication, but intensity modulation with direct detection (IM-DD) is most commonly used method. The expression for bit-error-rate (BER) for performance evaluation of IM-DD with ON/OFF shift-keying model is calculated using Poisson model that is basically given by the number of photons reached at the photon counter. In order to compute the number of photons arrived at the node $n$ in the time duration $T$, can be expressed as follows
\begin{equation}\label{eq13}
\begin{aligned}
\rho = \frac{P_{O_{r_n}}\eta_{n}\lambda}{T{D_{r}}h\hat{c}}
\end{aligned},
\end{equation}
where $D_r$ is the data transmission rate, $h$ is the Planck's constant, and $\hat{c}$ is the respective speed of light in water. The BER for photons arriving at node $n$ can be computed as
\begin{equation}\label{eq14}
\begin{aligned}
g_{n} = \frac{1}{2} \text{erfc} \left[\sqrt{\frac{T}{2}}(\sqrt{d_{1}} -\sqrt{d_{0}})\right]
\end{aligned},
\end{equation}
where $d_{1}= d_{r}+d_{n}+d_{g}$ and $d_{0}= d_{r}+d_{g}$ symbolize the number of photons essential for binary transmission 0 and 1, respectively. $\text{erfc} (.)$ is the complementary error function for the photons arriving at node $n$. Also, $d_r$ is the  dark count noise, while $d_g$ refer to the noise added as result of contextual enlightenment. According to \cite{vavoulas2014underwater}, by substituting the values of $d_1$ and $d_0$ in \eqref{eq14} and solving for $d_n$ we get
\begin{equation}\label{eq15}
\begin{aligned}
d_{n} = \left( \sqrt{d_{r}+d_{g}} + \sqrt{\frac{2}{T}} \text{erfc}^{-1}(2g_{n})\right)^{2} - d_{r} - d_{g}
\end{aligned}
\end{equation}

Now, putting \eqref{eq13}, and \eqref{eq14} in \eqref{eq15}, and solving it yields the optical distance $\tilde{r}_{O_{mn}}$ between node $m$ and node $n$ as
\begin{equation}\label{eq16}
\begin{aligned}
\tilde{r}_{O_{mn}} = \frac{2\cos{\theta}}{l(\lambda)} W_{0} \left[ \frac{l(\lambda)}{2\cos{\theta}\sqrt{\frac{2\pi Th\hat{c}D_{r}d_{n}(1-\cos{\theta_{0}})}{\eta_{n}\lambda P_{O_{t_m}}\eta_{m}\eta_{n}A_{n}\cos{\theta}}}}\right]
\end{aligned},
\end{equation}
where $W_{0} (\cdot)$ is known as the real part of Lambert $W_{0}$ function. 

\subsection{Hybrid MIAO Ranging Model}\label{subsec3.5}
 The noisy range measurements $\tilde{r}_{MI_{mn}}$, $\tilde{r}_{A_{mn}}$, and $\tilde{r}_{O_{mn}}$, are calculated using \eqref{eqMI3}, \eqref{eq9}, and \eqref{eq16}, respectively. In order to define the matrix for all the noisy range measurements, consider that
 \begin{equation}
 \tilde{r}_{mn} \approx \begin{cases}
 \tilde{r}_{MI_{mn}}\; \;\text{for MI link}, \\
 \tilde{r}_{A_{mn}}\;\;\;\;\text{for acoustic link},\\
 \tilde{r}_{O_{mn}} \; \;\;\;\text{for optical link}.
 \end{cases}
 \end{equation}

Now combining them all together we get
\begin{equation}\label{eq17}
\begin{aligned}
\boldsymbol{\aleph} = \left\{\tilde{r}_{mn} \right\}_{m,n=1, m \neq n}^{K} 
\end{aligned},
\end{equation}
where $K=M+N$ represents the total number of anchor and sensor nodes. In order to formulate the define problem for total number of nodes $K$, all dimensional spaces $\tilde{r}_{mn}$ are well approximated as ${r}_{mn}$. A cost function can be defined as follows which minimizes the error between the actual range and measured range 
\begin{equation}\label{eq18}
\begin{aligned}
\text{min}_{L} \sum_{m<n} \gamma_{mn} \left({\tilde{r}_{mn}} - r_{mn}(\textbf{\textit{L}})\right)^{2},
\end{aligned}
\end{equation}
where $\textbf{\textit{L}}=\{\textbf{\textit{l}}_{1},\textbf{\textit{l}}_{2},\cdots,\textbf{\textit{l}}_{K}\}$ are the respective 3D coordinates of all $K$ number of nodes. Also, $\gamma_{mn}$ is the weight between any two nodes $m$ and $n$, and is defined based on the ranging error variance, i.e., $\gamma_{mn}=\frac{1}{\sigma_{mn}^2}$, where the noisy range measurements are down-weighted by the large noise variance and vice-versa. Note that If there is no range measurement available between $m$-th and $n$-th node, then $\gamma_{mn} = 0$. Also, we assume that   ${\textstyle\sum}\gamma_{mn}=1$, $\gamma_{mn} \geq 0$, $\gamma_{mm} = 0$, and $\gamma_{mn} = \gamma_{nm}$, i.e., symmetric weights. In literature, many techniques have been presented to tackle the optimization problem, but none of those techniques uses more than one ranging measurement. Here, we propose a hybrid MIAO model that utilizes multiple input ranging measurements. For the hybrid MIAO ranging, the objective function is characterized as follows
\begin{equation}\label{eq19}
\begin{aligned}
\text{min}_{\eth^{(w)},L} \sum_{w=1}^{D} \eth^{(w)^{\tau}}  \sum_{m<n} \gamma_{mn} \left({\tilde{r}_{mn}} - r_{mn}(L)\right)^{2}
\end{aligned},
\end{equation}
where $\eth^{(w)}$  shows the importance of $w$-th iteration such that $\sum_{w=1}^{D}\eth^{(w)}=1$. Also, the term $\tau$ is the controlling factor and $D$ represent the total number of dimensions. The controlling factor for weighting is assumed to be $(\tau >1)$, in order to define the overall distribution of multiple observations.  Selecting only one view is not optimum in any case while ignoring the rest. Therefore, the proposed model assumes the combination of weights for each iteration view. 

\section{Proposed Localization Technique}\label{sec4}
Utilizing the BMDS technique and Procrustes analysis, the proposed method is characterized as follows:
\begin{enumerate} [(a)]
\item The MIAO based single hop ranging is used to find the shortest path distances among all pair of anchor and sensor nodes in the given network. Once all the pairwise distances are obtained, the completion of distance information matrix $\textbf{R} =\{\rho_{mn}^{2}\}_{m,n=1}^{K}$ is achieved, i.e., 
\begin{equation}\label{eq20}
  \textbf{R} =
  \begin{bmatrix}
    0 & \ldots & \rho_{1K}^{2} \\
    \vdots & \ddots & \vdots \\
     \rho_{K1}^{2} & \ldots & 0
  \end{bmatrix}.
\end{equation}
Matrix $\textbf{R}$ is a square symmetric matrix. 

\item BMDS technique is applied to matrix  $\textbf{R}$ for the estimation of relative coordinates. According to \cite{r31}, Kruskal outlined a stress function that basically minimizes the loss function to estimate the relative coordinates, i.e.,
\begin{equation}\label{eq21}
\zeta(\rho_{mn} | \textbf{Y}) = \frac{\sqrt{\sum_{m \neq n=1,\cdots,K}(\rho_{mn}-r_{mn})^{2}}}{{\sum_{m \neq n=1,\cdots,K}(\rho_{mn})^{2}}}
\end{equation}
Minimizing the above loss functions yields the estimated positions of all the nodes that is expressed as $\hat{}\textbf{Y} = \{{\alpha}_{m}\}_{m=1}^{K}$.  The minimization is achieved by first double centering matrix $\textbf{R}$, i.e.,
\begin{equation}\label{eq22}
\mathbf{\Gamma} = - {\frac{1}{2}} (\textbf{ARA}),
\end{equation}
where $\textbf{A} = \textbf{I}_{K} - (\frac{1}{K})\textbf{1}_{K} \textbf{1}_{K}^{T}$ is termed as operator for centering, where as $\textbf{I}_{K}$ is a $K \times K$ identity matrix. Also, $\textbf{1}_{K}$ is considered as the vector of ones with size of $k$. Eigen value decomposition of $\textbf{R}$ yields
\begin{equation}\label{eq23}
\text{EV} (\mathbf{\Gamma}) = \textbf{u}\textbf{v} \textbf{u}^{T}.
\end{equation}
The relative coordinates of all available nodes are measured utilizing the two largest eigenvalues provided by $\textbf{v}$, and the corresponding eigenvectors $\textbf{u}$ as follows
\begin{equation}\label{eq24}
\hat{\textbf{Y}} = \textbf{u} \sqrt{\textbf{v}},
\end{equation}
which can also be written as
\begin{equation}\label{eq25}
\hat{\textbf{Y}} = \{ \hat{\alpha}_{m}\}_{m=1}^{K}.
\end{equation}
The terms $\hat{\alpha}_{m}=\{\hat{x}_{m},\hat{y}_{m}, \hat{z}_{m}\}$ depicts the relative coordinates of node $m$ in the given network. Note that $\hat{\textbf{Y}}$ consists of relative coordinates of both the anchor nodes and sensor nodes. This is because the difference between the actual and estimated location of anchors is used to transform the map generated via BMDS in the next stage. This way, we can find out the values of optimal translation, rotation, and scaling factors that are used to find out the global coordinates of other sensor nodes. The translation, rotation, and scaling factors can be estimated using various linear transformation techniques, such as Procrustes analysis and principal component analysis.

\item Once BMDS calculates the initial estimated coordinates of the nodes. Then global transformation method is utilized to transform these initial estimated coordinates to their absolute equivalent coordinates. This method essentially finds out different parameters such as the scaling, rotation and translation. These parameters should be able enough to best maps the actual underlying coordinates. Assume that there exist $M$ number of anchor nodes in a 3D space, with actual coordinates equal to  $\alpha_{j} = \{x_{j},y_{j}, z_{j}\}$, and lies in range of $1\leq j \leq M$. It should be noted that the relative coordinates $\hat{\alpha}_{j} = \{\hat{x}_{j},\hat{y}_{j},\hat{z}_{j}\}$ must match the actual coordinates $\alpha_{j} = \{x_{j},y_{j},z_{j}\}$ for all the $M$ anchor nodes. Also, the relationship between the actual and relative coordinates is computed by
\begin{equation}\label{eq26}
\Upsilon = \sum_{j=1}^{M} {(\hat{\alpha}_{j} - \alpha}_{j})^{T}(\hat{\alpha}_{j} - \alpha_{j}).
\end{equation}
This relationship can also be expressed in terms of an objective function  $q(\xi,\nu,\kappa)$, where $\xi$ is the scaling factor, $\nu$ is the rotation factor, and $\kappa$ is the respective translational factor. According to linear transformation, the objective function can be written as 
\begin{equation}\label{eq27}
q(\xi,\nu,\kappa) = \sum_{j=1}^{M} {(\hat{\alpha}_{j} - \xi \kappa^{T}\alpha_{j}} -\nu)^{T} \times (\hat{\alpha}_{j} -\xi \kappa^{T}\alpha_{j} - \nu),
\end{equation}
The objective function in \eqref{eq27} can be minimizes by determining the optimum values of  $\xi$, $\nu$, and $\kappa$, i.e.,
\begin{equation}\label{eq28}
\{\bar{\xi}, \bar{\nu}, \bar{\kappa}\} = \arg_{\xi, \nu, \kappa} \text{min} \hspace{0.025in} q (\xi, \nu, \kappa).
\end{equation}
For anchor nodes, assuming that $\textbf{b}_{0}$ and $\textbf{c}_{0}$ are the centroids  for the actual and estimated locations, respectively. These centroids are represented as
\begin{equation}\label{eq29}
\textbf{b}_{0} = \frac{1}{M} \sum_{j=1}^{M} \alpha_{j},
\end{equation}
and
\begin{equation}\label{eq30}
\textbf{c}_{0} = \frac{1}{M} \sum_{j=1}^{M} \hat{\alpha}_{j}.
\end{equation}
In order to achieve optimum translation, the objective function is expressed as
\begin{equation}\label{eq31}
\begin{aligned}
q(\xi,\nu,\kappa) = \hspace{2.5in} \\ \sum_{j=1}^{M} {\bigg((\hat{\alpha}_{j} - \textbf{c}_{0}) - \xi \kappa^{T}(\alpha_{j} - \textbf{b}_{0})} + \hat{\alpha}_{j} - \xi \kappa^{T}\alpha_{j} - \nu \bigg)^{T}  \\
\times  \bigg((\hat{\alpha}_{j}-\textbf{c}_{0} )-\xi\kappa^{T}(\alpha_{j}-\textbf{b}_{0}) + 
 \hat{\alpha}_{j} - \xi \kappa^{T}\alpha_{j} - \nu \bigg)
\end{aligned}
\end{equation}
After seprating the term $\hat{\alpha}_{j} - \xi \kappa^{T}\alpha_{j} - \nu$ from equation \eqref{eq31}, we get
\begin{equation}\label{eq32}
\begin{aligned}
q(\xi,\nu,\kappa) = \sum_{j=1}^{M} \Bigg\{ {\bigg((\hat{\alpha}_{j} - \textbf{c}_{0}) - \xi \kappa^{T}(\alpha_{j} - \textbf{b}_{0})\bigg)^{T}} \\ 
\times \bigg((\hat{\alpha}_{j} - \textbf{c}_{0}) -
 \xi \kappa^{T}(\alpha_{j} - \textbf{b}_{0}) \bigg) \hspace{0.45in} \\ 
+ M(\hat{\alpha}_{j}-\xi\kappa^{T}\alpha_{j} - \nu)^{T}
(\hat{\alpha}_{j} - \xi \kappa^{T}\alpha{j} - \nu) \Bigg\}.
 \end{aligned}
\end{equation}
Now using \eqref{eq32}, the optimum translation factor $\overline{\nu}$ that minimizes the above objective function, can be expressed as
\begin{equation}\label{eqnew}
\overline{\nu} = (\hat{\alpha}_{j} - \overline{\xi}\overline{\kappa}^{T}\alpha_{j}).
\end{equation}
In terms of the centroids $\textbf{c}_{0}$ and $\textbf{b}_{0}$, \eqref{eqnew} can be written as
\begin{equation}\label{eq33}
\overline{\nu} = \textbf{c}_{0} - \overline{\xi}\overline{\kappa}^{T}\textbf{b}_{0}.
\end{equation}
By considering the condition of $\textbf{b}_{0} = \textbf{c}_{0} = 0$ and substituting \eqref{eq33} into \eqref{eq32}, the objective function can be simplified as follows
\begin{equation}\label{eq34}
\begin{aligned}
q(\xi,\nu,\kappa) = \sum_{j=1}^{M} {(\hat{\alpha}_{j} -\xi\kappa^{T}\alpha_{j})^{T} (\hat{\alpha}_{j} -\xi\kappa^{T}\alpha_{j})}.
 \end{aligned}
\end{equation}

The above objective function obtained is convex and differentiating it with respect to $\xi$, the optimum value $\overline{\xi}$ that minimizes the overall function $q(\xi, \nu, \kappa)$ and is calculated as
\begin{equation}\label{eq35}
\overline{\xi} = \frac{\text{Tr}(\alpha_{j}\overline{\kappa}\hat{\alpha}_{j}^{T})}{\text{Tr}(\alpha_{j}\hat{\alpha}_{j}^{T})},
\end{equation}
where  $\text{Tr}({\cdot})$ is known as the trace operator. The optimum rotation matrix $\overline{\kappa}$ is expressed in terms Eigen-decomposition of the factors  $\alpha_{j}\hat{\alpha}_{j}$ as
\begin{equation}\label{eq36}
\overline{\kappa} = u {v}'
\end{equation}

whereas $u$ and ${v}'$ are the respective eigenvectors and eigenvalues of  $\alpha_{j}\hat{\alpha}_{j}$. Likewise, the term $\overline{\kappa}$ can be inscribed in terms of 
\begin{equation}\label{eq37}
\overline{\kappa} = \frac{\sqrt{(\alpha_{j}^{T}\hat{\alpha}_{j}\hat{\alpha}_{j}^{T}\alpha_{j}})} {(\hat{\alpha}_{j}^{T}\alpha_{j})}.
\end{equation}
Finally, with the use of optimum parameters $\overline{\xi}$, $\overline{\nu}$, $\overline{\kappa}$, the actual positions of all nodes in the given network can easily to be computed by

\begin{equation}\label{eq38}
\tilde{\textbf{Y}}= \overline{\xi}\overline{\kappa}^{T}({\hat{\textbf{Y}})} + \overline{\nu}.
\end{equation}

  \item{\textbf{Energy Consumption Vs. Localization Accuracy:}} 
The IoUT network comprises battery-operated sensor nodes with a limited amount of onboard energy operating in a harsh environment. Therefore, these IoUT networks require the designing of energy-efficient protocols to improve the lifetime of the network.  In the case of our proposed MIAO model, we can express the total energy consumed by all $M+N$ nodes as
\begin{equation}
\begin{aligned}\label{eq:ET}
\mathcal{E}_{T}= \sum_{m=1}^{M+N} E_{F_{m}} + (M+N)\sum_{m=1}^{M+N} E_{R_{m}},
\end{aligned}
\end{equation}
where $E_{F}=\sum_{m=1}^{M+N} E_{F_{m}}$ is the $m$-th node system's electronic circuitry for fundamental operation, and $E_{R}=(M+N)\sum_{m=1}^{M+N} E_{R_{m}}$ represents the energy consumed during the transmission.
The $E_{R_m}$ for the $m$-th node can be written as
\begin{equation}
\begin{aligned}\label{eq:ER}
E_{R_m}= E_{B} \left(\frac{4\pi R_m}{\lambda}\right)^{2},
\end{aligned}
\end{equation}
where $E_{B}$ is the energy required for a single bit transmission and $R_m$ is the achievable transmission range of $m$-th node.  It is clear from \eqref{eq:ER}, as energy consumption is proportional to the square of the transmission range for each node in the given hybrid network. 
Both communication energy consumption and localization accuracy rely on the transmission range $R_m$. Therefore, we introduce the following energy error product as a benchmark to confer the trade-off between energy consumption and localization accuracy
\begin{equation}
\begin{aligned}\label{eq:CR}
C(R)= E_{R} \times \text{(RMSE)},
\end{aligned}
\end{equation}
where RMSE represents the average root-mean-square-error of all nodes in the given network. The RMSE as a function of estimated and actual nodes locations can be written as
  \begin{equation}
\begin{aligned}\label{eq:RMSE}
\text{RMSE} = \hspace{2.5in} \\
\sum_{m=1}^{M+N} \frac{\sqrt{(\hat{x}_{m} - x_{m})^{2} + (\hat{y}_{m} - y_{m})^{2} + (\hat{z}_{m} - z_{m})^{2}}}{M+N}.
\end{aligned}
\end{equation}
Now substituting \eqref{eq:ER} and \eqref{eq:RMSE} in \eqref{eq:CR} results in
\begin{equation}
\begin{aligned}\label{eq:CRRMSEN}
C(R)= E_{B} \left(\frac{4\pi}{\lambda}\right)^{2} \sum_{m=1}^{M+N} R_{m}^{2} \hspace{1.35in} \\
\times \sum_{m=1}^{M+N}  \frac{\sqrt{(\hat{x}_{m} - x_{m})^{2} + (\hat{y}_{m} - y_{m})^{2} + (\hat{z}_{m} - z_{m})^{2}}}{M+N}.
\end{aligned}
\end{equation}
Simplifying \eqref{eq:CRRMSEN} yields
\begin{equation}
\begin{aligned}\label{eq:CRRMSEE}
C(R)= E_{B} \left(\frac{4\pi}{\lambda \sqrt{M+N}}\right)^{2} \hspace{1.5in} \\
\times \sum_{m=1}^{M+N} R_{m}^{2} \sqrt{(\hat{x}_{m} - x_{m})^{2} + (\hat{y}_{m} - y_{m})^{2} + (\hat{z}_{m} - z_{m})^{2}}.
\end{aligned}
\end{equation}\par

According to \eqref{eq:CRRMSEE}, the energy-error product can be optimized by finding the minimum of $C(R)$ function.

 \item {\textbf{Computational Complexity:}}
  The fundamental step in finding a sensor node's location in network localization schemes is measuring the pairwise distances. Afterward, an optimization method needs to be followed to minimize the inconsistency between the estimated pairwise distances and actual Euclidean distances. Usually, single-hop distances among the neighbors are measured using ranging methods \cite{schwartz1989numerical, tenenbaum2000global}. 
The complexity to estimate the pairwise distances from these single-hop distances is $C(K^{3})$, where $K=M+N$ (total number of nodes). Moreover, the computational complexity for global transformation, i.e., from local coordinates to global using anchor nodes is $C(M^{2}) + C(K)$. Therefore, the total time complexity can be expressed as
  
  \begin{equation}\label{complexity1}
  \begin{aligned}
  \text{Complexity} = C(K^{3}) + C(M^{2}) + C(K),
  \end{aligned}
  \end{equation}

\noindent in \eqref{complexity1}, the term $C(K^{3}$) represents the total complexity which is dominated by $C(K^{3})$.

\end{enumerate}

\section{Performance Evaluation of the Proposed Model}\label{sec5}
Gaussian noise usually effect the performance of the ranging measurements $\rho_{mn}$. As, the noise is probabilistic in nature, hence affecting the overall ranging measurements. Therefore, the probability density function (PDF) of the ranging measurements $\rho_{mn}$ with specified locations of the anchor node $m$ and sensor node $n$ can be expressed by $s(\rho_{mn} | \alpha_{m}, \alpha_{n})$ in terms of
\begin{equation}\label{eq39}
s(\rho_{mn} | \alpha_{m}, \alpha_{n}) = \frac{1}{\psi_{d} \sqrt{2\pi}} \exp^{\left({-\frac{(\rho_{mn}-r_{mn})^{2}}{2\psi_{d}^{2}}}\right)}.
\end{equation}
It should be noted that variance of the calculated measurements for range estimation are correlated to the distance by $\psi_{d}^{2}=\varepsilon r_{mn}^{\delta_{mn-1}}$. The consequent ratio of log likelihood is depicted as $E_{mn} = \log(s(\rho_{mn}| \alpha_{m}, \alpha_{n}))$. The same when converted to decibel (dB) scale can be written as follows
\begin{equation}\label{eq40}
\begin{aligned}
E_{mn} \text{[dB]} = - \text{log} \sqrt{2\pi\varepsilon} - \frac {\delta_{mn}}{4} \text{log}({\vert\vert} \alpha_{m} - \alpha_{n} {\vert \vert}^{2}) \\
- \frac{1}{2\varepsilon} \frac{(\rho_{mn}-{\vert\vert}\alpha_{m} - \alpha_{n}{\vert\vert})^{2}}{{{(\vert\vert}\alpha_{m} - \alpha_{n}{\vert\vert}^{2})}^{\frac{\delta_{mn}}{2}}}
\end{aligned}
\end{equation}
It is further anticipated that $\rho_{mn}$ is degraded by the environmental noise, independent in nature. So, considering the effect of noise with joint ratio of log-likelihood to estimate the range of almost each set of pairwise measurements.
\begin{equation}\label{eq41}
\begin{aligned}
\mathbf{\Psi}_{mn} = \sum_{m=1}^{M} \sum_{n=m+1}^{M+N} \log (s(\rho_{mn} | \alpha_{m}, \alpha_{n}))
\end{aligned},
\end{equation}
where $M+N$ shows the total number of anchor and sensor nodes. Further solving \eqref{eq41} yields
\begin{equation}\label{eq42}
\begin{aligned}
\mathbf{\Psi}_{mn} = \sum_{m=1}^{M} \sum_{n=m+1}^{M+N} \{ - \text{log} \sqrt{2\pi\varepsilon} -\frac{\delta_{mn}}{4}\text{log}{({\vert\vert}\alpha_{m} - \alpha_{n}{\vert\vert}^{2})} \\
- \frac{1}{2\varepsilon} \frac{(\rho_{mn}-{\vert\vert}\alpha_{m} - \alpha_{n}{\vert\vert})^{2}}{{{(\vert\vert}\alpha_{m} - \alpha_{n}{\vert\vert}^{2})}^{\frac{\psi_{mn}}{2}}} \}
\end{aligned}
\end{equation}

Based on this log-likelihood ratio, the Hybrid Cramer-Rao lower bound (H-CRLB) is derived. The H-CRLB is basically a lower threshold on the estimator variance that is unbiased \cite{r32}, and thus, providing a benchmark to accurately evaluate the designed algorithms in terms of performance. The H-CRLB technique is  based on Fisher Information Matrix (FIM) \cite{r33} represented by $\mathbf{\Phi}$ and expressed as
\begin{equation}\label{eq43}
\begin{aligned}
\mathbf{\Phi}= - \mathbb{C}_{\alpha_{m},\alpha_{n}} (\Delta_{\alpha_{m},\alpha_{n}} (\Delta_{\alpha_{m},\alpha_{n}}(\mathbf{\Psi}_{mn})))
\end{aligned},
\end{equation}
 where $\mathbb{C}_{\alpha_{m},\alpha_{n}}$ is  termed as the expected value for second order derivative of the ratio of log-likelihood $\mathbf{\Psi}_{mn}$. The FIM can further be inscribed in terms of sub-matrices as follows
\begin{equation}\label{eq44}
  \mathbf{\Phi} =
  \begin{bmatrix}
    \mathbf{\Phi}_{xx} & \mathbf{\Phi}_{xy} & \mathbf{\Phi}_{xz}  \\
    \mathbf{\Phi}_{xy}^{T} & \mathbf{\Phi}_{yy} & \mathbf{\Phi}_{yz}  \\
    \mathbf{\Phi}_{xz}^{T} & \mathbf{\Phi}_{yz}^{T} & \mathbf{\Phi}_{zz}  \\
     \end{bmatrix}
\end{equation}
The subscripts $xx$, $yy$ and $zz$ denotes the diagonal sub-matrices for the considered FIM. On the other hand, the $xy$, $xz$, $yz$, ${xy}^T$, ${xz}^{T}$ and ${yz}^T$ denotes the sub-matrices that are non-diagonal. Furthermore, the diagonal elements for each sub-matrix ranges from $m=1, 2, 3,\cdots, K$ are given as 
\begin{equation}\label{eq45}
\begin{aligned}
\mathbf{\Phi}_{xx(m,m)} = \sum_{n \in H(m)} \frac{1}{\psi_{d}^{2}} \frac{\beta_{mn}(x_{m} - x_{n})^{2}}{{\vert\vert}\alpha_{m} - \alpha_{n}{\vert\vert}^{2}}
\end{aligned},
\end{equation}
\begin{equation}\label{eq46}
\begin{aligned}
\mathbf{\Phi}_{xy(m,m)} = \sum_{n \in H(m)} \frac{1}{\psi_{d}^{2}} \frac{\beta_{mn}(x_{m} - x_{n})^{2}(y_{m} - y_{n})^{2}}{{\vert\vert}\alpha_{m} - \alpha_{n}{\vert\vert}^{2}}
\end{aligned},
\end{equation}
\begin{equation}\label{eq47}
\begin{aligned}
\mathbf{\Phi}_{yy(m,m)} = \sum_{n \in H(m)} \frac{1}{\psi_{d}^{2}} \frac{\beta_{mn}(y_{m} - y_{n})^{2}}{{\vert\vert}\alpha_{m} - \alpha_{n}{\vert\vert}^{2}}
\end{aligned},
\end{equation}
\begin{equation}\label{eq48}
\begin{aligned}
\mathbf{\Phi}_{yz(m,m)} = \sum_{n \in H(m)} \frac{1}{\psi_{d}^{2}} \frac{\beta_{mn}(y_{m} - y_{n})^{2}(z_{m} - z_{n})^{2}}{{\vert\vert}\alpha_{m} - \alpha_{n}{\vert\vert}^{2}}
\end{aligned},
\end{equation}
\begin{equation}\label{eq49}
\begin{aligned}
\mathbf{\Phi}_{xz(m,m)} = \sum_{n \in H(m)} \frac{1}{\psi_{d}^{2}} \frac{\beta_{mn}(x_{m} - x_{n})^{2}(z_{m} - z_{n})^{2}}{{\vert\vert}\alpha_{m} - \alpha_{n}{\vert\vert}^{2}}
\end{aligned},
\end{equation}
and
\begin{equation}\label{eq50}
\begin{aligned}
\mathbf{\Phi}_{zz(m,m)} = \sum_{n \in H(m)} \frac{1}{\psi_{d}^{2}} \frac{\beta_{mn}(z_{m} - z_{n})^{2}}{{\vert\vert}\alpha_{m} - \alpha_{n}{\vert\vert}^{2}}
\end{aligned},
\end{equation}
respectively. Similarly, the non-diagonal sub-matrices, where $m \neq n$ for $m,n=1,2,\cdots,K$ are given as
\begin{equation}\label{eq51}
\begin{aligned}
\mathbf{\Phi}_{xx(m,n)} = \frac{-1}{\psi_{d}^{2}} \frac{\beta_{mn}(x_{m} - x_{n})^{2}}{{\vert\vert}\alpha_{m} - \alpha_{n}{\vert\vert}^{2}}
\end{aligned},
\end{equation}
\begin{equation}\label{eq52}
\begin{aligned}
\mathbf{\Phi}_{xy(m,n)} = \frac{-1}{\psi_{d}^{2}} \frac{\beta_{mn}(x_{m} - x_{n})^{2}(y_{m} - y_{n})^{2}}{{\vert\vert}\alpha_{m} - \alpha_{n}{\vert\vert}^{2}}
\end{aligned},
\end{equation}
\begin{equation}\label{eq53}
\begin{aligned}
\mathbf{\Phi}_{yy(m,n)} = \frac{-1}{\psi_{d}^{2}} \frac{\beta_{mn}(y_{m} - y_{n})^{2}}{{\vert\vert}\alpha_{m} - \alpha_{n}{\vert\vert}^{2}}
\end{aligned},
\end{equation}
\begin{equation}\label{eq54}
\begin{aligned}
\mathbf{\Phi}_{yz(m,n)} = \frac{-1}{\psi_{d}^{2}} \frac{\beta_{mn}(y_{m} - y_{n})^{2}(z_{m} - z_{n})^{2}}{{\vert\vert}\alpha_{m} - \alpha_{n}{\vert\vert}^{2}}
\end{aligned},
\end{equation}
\begin{equation}\label{eq55}
\begin{aligned}
\mathbf{\Phi}_{xz(m,n)} = \frac{-1}{\psi_{d}^{2}} \frac{\beta_{mn}(x_{m} - x_{n})^{2}(z_{m} - z_{n})^{2}}{{\vert\vert}\alpha_{m} - \alpha_{n}{\vert\vert}^{2}}
\end{aligned},
\end{equation}
and 
\begin{equation}\label{eq56}
\begin{aligned}
\mathbf{\Phi}_{zz(m,n)} = \frac{-1}{\psi_{d}^{2}} \frac{\beta_{mn}(z_{m} - z_{n})^{2}}{{\vert\vert}\alpha_{m} - \alpha_{n}{\vert\vert}^{2}}
\end{aligned},
\end{equation}

The term $\beta_{mn}$ in the above expressions is known as the scaling factor that is dependent on the distance estimated among the nodes and can be written as 
\begin{equation}\label{eq57}
\begin{aligned}
\beta_{mn} = 1 + \frac{\delta_{mn}^{2}\varepsilon}{2}{r_{mn}^{\delta_{mn}-2}}
\end{aligned},
\end{equation}
On the basis of $\mathbf{\Phi}$, the H-CRLB for position estimation of the sensor nodes can be specifically expressed as
\begin{equation}\label{eq58}
\begin{aligned}
\text{H-CRLB} = \text{Tr}\{\mathbf{\Phi}^{-1} \}
\end{aligned},
\end{equation}
The term $\mathbf{\Phi}^{-1}$ is  the inverse of FIM. The relationship between the Root-Mean-Square-Error (RMSE) (given by \eqref{eq:RMSE})and derived H-CRLB is given as
\begin{equation}\label{eq59}
\begin{aligned}
 \text{RMSE} \geq \text{H-CRLB}
\end{aligned}.
\end{equation}

\section{Simulation Results and Discussion}\label{sec6}
MATLAB is used as a simulation tool for performance evaluation of the proposed MIAO ranging and localization technique for a given IoUT network. The parameters used for the simulations are summarized in Table \ref{tab:table1}.

\begin{table}[ht]
\caption{Parameters used for simulations.}\label{tab:table1}
\centering
\begin{tabular}{|p{4.5cm}|p{3.5cm}| }
\hline
\hline
\textbf{Parameters} & \textbf{Values}  \\    
\hline
\hline
Simulation Area & 100m $\times$ 100m $\times$ 100m \\
\hline
Permeability $\mu$ & 4 $\pi \times$ 10$^{-7}$ H/m \\
\hline
Ocean water conductivity $\sigma$ & 4 S/m \\
\hline
Clean water conductivity $\sigma$ & 0.01 S/m \\
\hline
Noise Power $P_{n}$ & $2 \times 10^{-6}$ W \\
\hline
Noise variance & 0-1 m \\
\hline
Number of anchor nodes $M$ & 4 to 20 \\
\hline
Number of sensor nodes $N$ & 96 to 150 \\
\hline
Number of relay nodes & 4 \\
\hline 
\hline     
\end{tabular}
\end{table} 

  \subsection{MIAO-based position estimation}\label{subsec6.1}
\begin{figure}[htp!]
\centering
\begin{center}
\includegraphics[width=1.06\columnwidth]{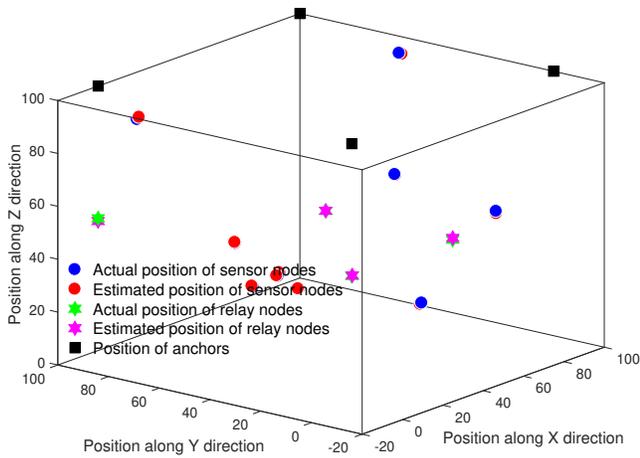}  
\caption{MIAO-based position estimation of sparsely deployed nodes in a cubic area of 100m $\times$ 100m $\times$100m.\label{scenario1}}  
\end{center}  
\end{figure}

\begin{figure}[htp!]
\begin{center}
\includegraphics[width=1.02\columnwidth]{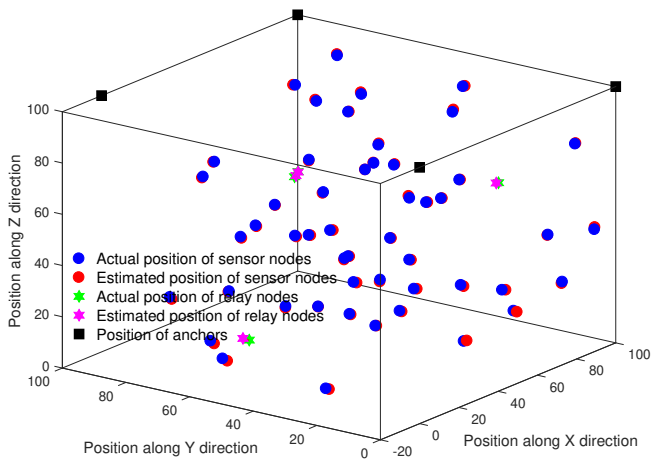}  
\caption{MIAO-based position estimation of densely deployed nodes in a cubic area of 100m $\times$ 100m $\times$100m.\label{scenario2}}  
\end{center}  
\end{figure}

We mainly consider the RMSE as a performance metric to test various parameters of the proposed model. Utilizing \eqref{eq:RMSE}, where \{$\hat{x}_{m},\hat{y}_{m},\hat{z}_{m}$\} and \{$x_{m}, y_{m}, z_{m}$\} are the respective estimated and actual coordinates of the anchor node $m$. We consider a 3D IoUT network setup with sensor nodes, anchors nodes, and relay nodes. Based on recent literature \cite{chen2021optimizing,liu2020advances,xing2020joint}, it is efficient to consider relay nodes in three-dimensional underwater networks. Therefore, in our simulation setup, we considered four relay nodes. The relay nodes can be AUVs or suspended sensors to collect the data from other sensor nodes efficiently. For performance analysis of the proposed method, we have considered two scenarios, as shown in Fig. \ref{scenario1} and \ref{scenario2}. The first scenario consists of sparsely distributed nodes in a 100m $\times$100m $\times$100m cubic area with four anchor nodes. In contrast, the second scenario comprises densely populated nodes in the same observation area. Both figures depict that increasing the number of nodes improves the location awareness accuracy mainly due to the low shortest path estimation error in dense networks. It should be noted that simulations are restricted to 100m $\times$ 100m $\times$ 100m cubic area. This is mainly due to the centralized nature of the algorithm, which requires a connected network. If we increase the network area with the same number of nodes, it will result in a disconnected graph, and Bayesian multidimensional scaling (BMDS) fails to operate on disconnected graphs.

\subsection{RMSE Vs. Ranging noise variance}\label{subsec6.2}
\begin{figure}[htp!]
\begin{center}
\includegraphics[width=0.925\columnwidth]{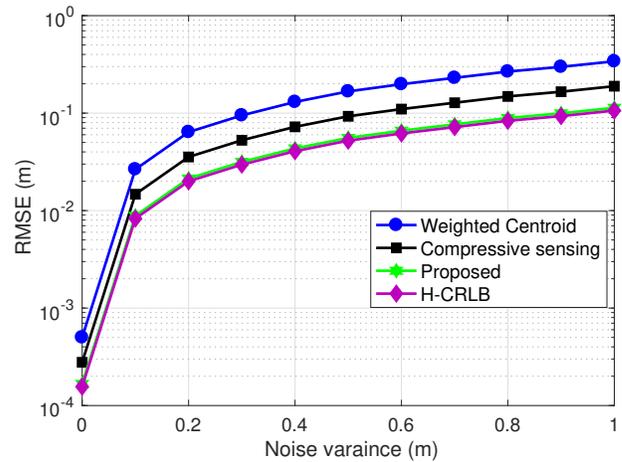}  
\caption{RMSE vs. Ranging noise variance.\label{noise_var}}  
\end{center}  
\end{figure}

We carried out the comparative analysis of the proposed method in terms of various parameters such as noise variance, network density, and number of anchors. Also, the results are compared to some well-known network localization techniques such as compressive sensing and weighted centroid localization (WCL) \cite{r34}. Intuitively, it is observed that the RMSE increases with an increase in the noise variance, as shown in Fig. \ref{noise_var}. Indeed, the estimation of missing distances and noise variance affects the overall accuracy of location awareness in the hybrid IoUT network. We have examined the results for proposed MIAO model in the presence of Gaussian noise distribution with zero mean and variance. For simulation results, values of the noise variance are set to approximately 0-1m. Further, it has been noted that the proposed model is robust to error variance as compared to WCL and compressive sensing. The reason behind is better distance approximation of the pair-wise missing distances. 

The ranging noise variance is proportionate to the accurate Euclidean distance $r_{mn}$ between the anchor node $m$ and sensor node $n$. Moreover, the proposed method outperforms the literature because the WCL is center-biased in nature and is not robust to noise. Similarly, the compressive sensing technique requires signal reconstruction from raw data and gets affected by the noisy underwater environment. Hence, the proposed method outperforms both WCL and compressive sensing techniques and is approximately equal to the H-CRLB, as shown in Fig. \ref{noise_var}. The improvement in the proposed method is because of the availability of the accurately estimated missing distances among the various nodes.  \par 

\subsection{RMSE Vs. Number of nodes}\label{subsec6.3}
\begin{figure}[htp!]
\begin{center}
\includegraphics[width=0.925\columnwidth]{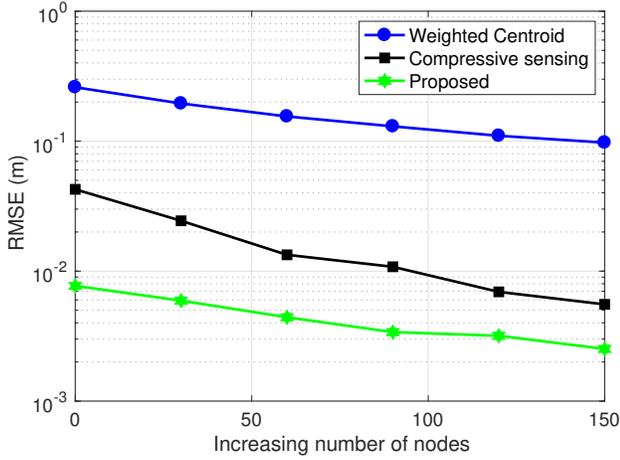}  
\caption{RMSE vs. Increasing number of nodes.\label{nodes}}  
\end{center}  
\end{figure}

To show the impact of the number of nodes on the localization accuracy, we consider the same setup of 100 nodes distributed nodes in a 100m $\times$100m $\times$100m cubic area with four anchor nodes. Fig. \ref{nodes} shows that increasing the number of nodes for location awareness in hybrid IoUT network results in an improved RMSE. This is mainly due to the improved connectivity of the network. Increasing the number of nodes provides a better shortest path estimation of the range measurements in dense networks than sparsely populated networks. 

\begin{figure}[htp!]
\begin{center}
\includegraphics[width=0.925\columnwidth]{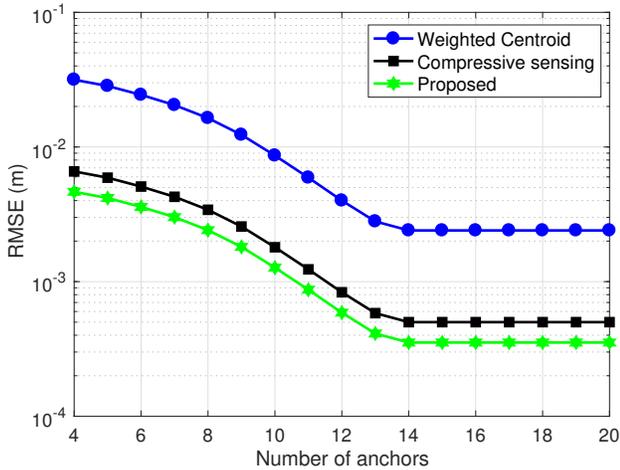}  
\caption{RMSE vs. Increasing number of anchors.\label{anchors}}  
\end{center}  
\end{figure}

For performance evaluation of the proposed hybrid technique in accordance with the number of anchor nodes, we have considered 96 sensor nodes and 4 anchor nodes in the same 100m $\times$ 100m $\times$ 100m cubic area. It can be observed from Fig. \ref{anchors}, that if the anchor nodes are increased up to 15, the hybrid IoUT network become saturated. No further increment in number of anchor nodes enhances the network localization capability, and hence, no improvement in RMSE. Therefore, it is important to model these parameters in a much more accurate way for the practical deployment of Hybrid-IoUT network.

  \subsection{RMSE Vs. Transmission range:}
\begin{figure}[htp!]
\centering
\begin{center}
\includegraphics[width=1.03\columnwidth]{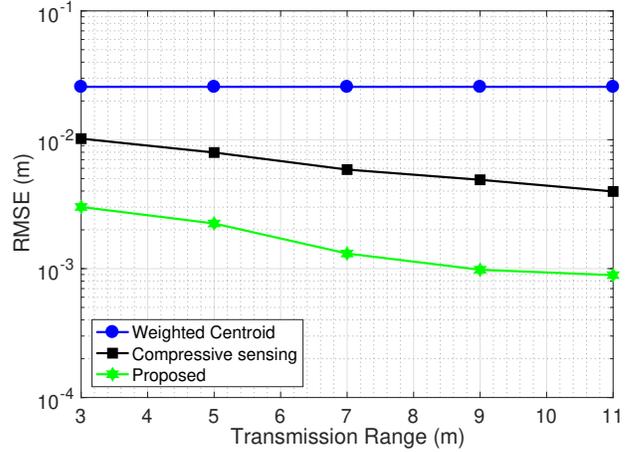}  
\caption{RMSE vs. Transmission Range.\label{energy211}}  
\end{center}  
    \end{figure}

We performed simulations to look for the relationship between RMSE and transmission range for the scenario of 100 nodes (96 sensor nodes and 4 relay nodes) deployed in a 100m $\times$100m $\times$100m cubic area. The anchor nodes are equal to four, and the noise variance is 0.1m. Fig.  \ref{energy211} shows that the RMSE decreases with an increase in transmission range $R$. As the node's transmission range increases, the network's average connectivity increases and decreases the RMSE up to a particular value, after which it saturates. For example, we can see that the  RMSE decreases with an increase in the transmission range up to 7m, after which it almost saturates, and further increase in transmission range will only increase the energy consumption.  
Fig.  \ref{energy211} also illustrates that the proposed approach outperforms WCL \cite{r34} and compressive sensing because both WCL and compressive sensing consider a fully connected network where an increase in the transmission range does not improve the localization accuracy.

\begin{figure}[htp!]
\centering
\begin{center}
\includegraphics[width=1\columnwidth]{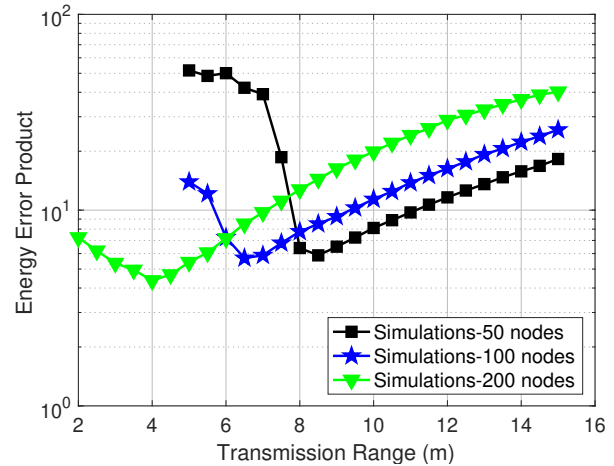}  
\caption{Energy error product vs. Transmission Range.\label{energy1}}  
\end{center}  
\end{figure}

Fig.  \ref{energy211} shows only the impact of transmission range on the localization accuracy; we further evaluate the energy error product as a function of the transmission range. For this,
we consider three different scenarios with 50, 100, and 200 IoUT nodes in 100m $\times$100m $\times$100m cubic area. Fig.  \ref{energy1} shows that the energy error product depends on both the number of nodes and the transmission range. 
For a dense network such as 200 nodes, the optimal transmission range with minimum energy error product is towards the lower end, i.e., 4m. For a sparse network, the minimum energy error product is around 8m. This is because increasing the transmission range improves the accuracy up to a specific value, after which the saturation occurs, and a further increase in range will only result in high energy consumption. Fig.  \ref{energy1} deduce that a larger transmission range reduces the localization error but leads to increased energy consumption. Hence, it is crucial to select a proper transmission range for IoUT nodes to maintain energy efficiency. Besides optimizing the above function, energy-harvesting techniques have recently attracted the researchers for improving the network lifetime of underwater communications systems \cite{saeed2018energy}. As the underwater nodes are not capable to survive on the battery-operation for larger time, thus, energy harvesting methods including microbial fuel cells \cite{srujana2015multi} and acoustic piezo-electric harvesters \cite{li2016energy} can be a promising solutions.

\section{Conclusion}
This paper introduces a hybrid MIAO ranging technique that utilizes a SOA approach for accurate localization.  Based on the SOA method, nearest-neighbour distances are measured. This method leads to better localization accuracy than a stand-alone single underwater communication technology due to the availability of many single-hop ranges in IoUT networks.  Moreover, Bayesian MDS is introduced to calculate the network graph from the computed single-hop distances. The output graph from the BMDS is fed to the Procrustes analysis technique to estimate the underwater IoUT devices' unknown location accurately.
Furthermore, the hybrid Cramer Rao lower bound is derived for analyzing the performance of the proposed method. Simulations are performed for sparse and dense IoUT networks to see the effectiveness of the proposed scheme. The results show the superior performance of the proposed method with respect to the literature in terms of different system parameters, such as ranging error variance, network density, and the total number of anchors. The simulation results depict that the proposed MIAO scheme achieves a sub-meter level of accuracy in even sparse IoUT networks.

In future work, we will perform in-situ measurements in a real underwater setup and compare the simulation and practical results both in terms of computational complexity, energy efficiency, and error performance.

\section*{Acknowledgement}
The authors would like thank the anonymous reviewers for their fruitful comments which further improved the scope of this manuscript.

\bibliographystyle{IEEEtran}
\bibliography{mybibliography}

\begin{IEEEbiography}[{\includegraphics[width=1in, height=1.25in, clip, keepaspectratio]{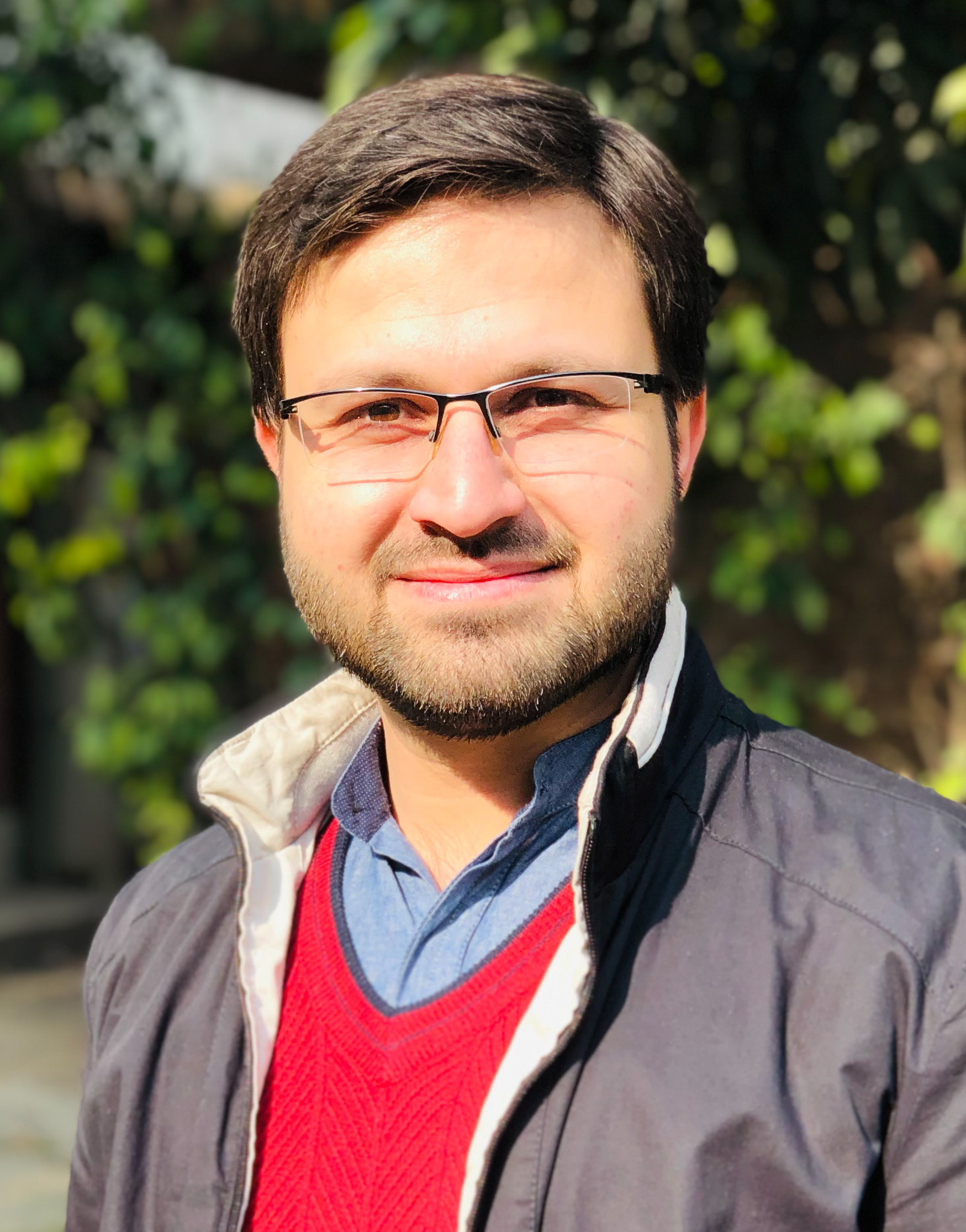}}]
{RUHUL AMIN KHALIL} (Member, IEEE) received the bachelor's, master's, and Ph.D. degrees in electrical engineering from the Department of Electrical Engineering, Faculty of Electrical and Computer Engineering, University of Engineering and Technology, Peshawar, Pakistan, in 2013, 2015, and 2021, respectively. He has been serving as a Lecturer with the Department of Electrical Engineering, Faculty of Electrical and Computer
Engineering, University of Engineering and Technology, Peshawar. His research interests include audio signal processing and its applications, machine learning, the Internet of Things (IoT), routing, network traffic estimation, software defined networks, and underwater wireless communication.
\end{IEEEbiography}

\begin{IEEEbiography}[{\includegraphics[width=1in, height=1.25in, clip, keepaspectratio]{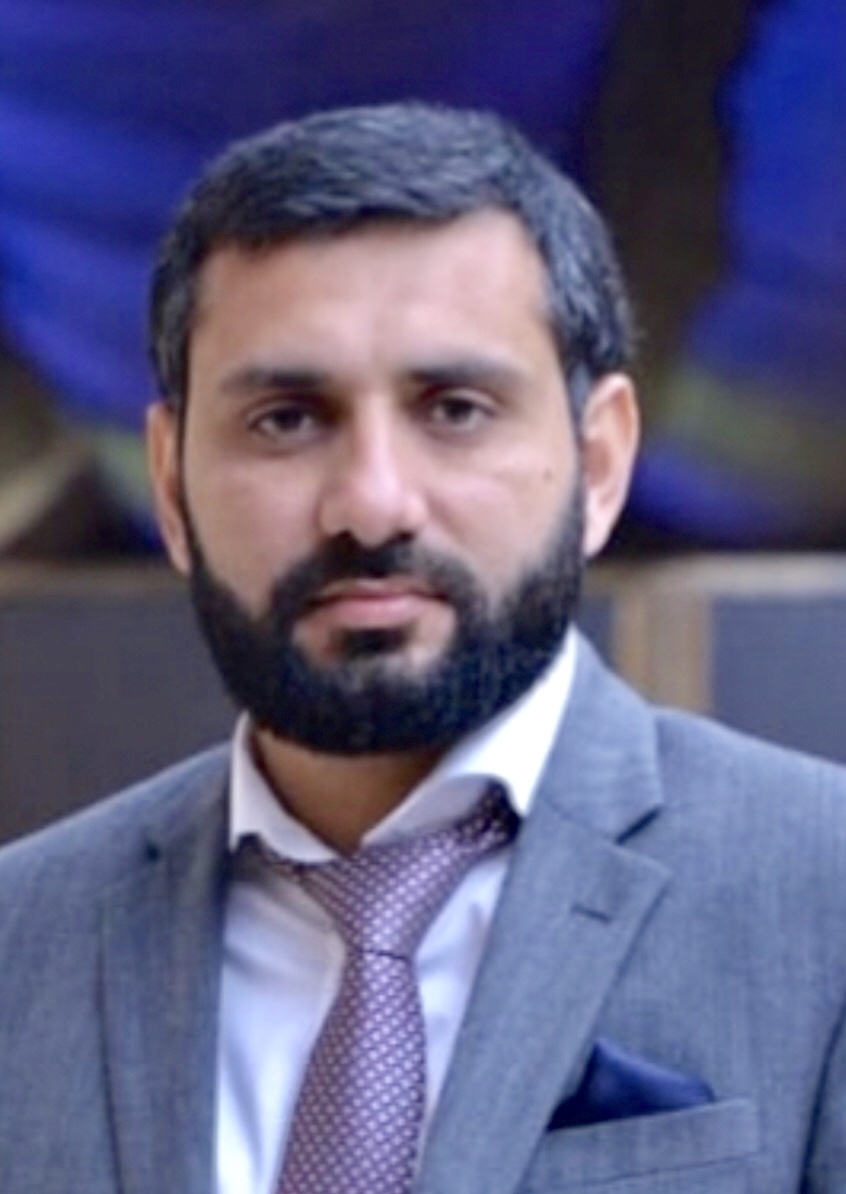}}]
{NASIR SAEED} (Senior Member, IEEE) received the bachelor's degree in telecommunication from the University of Engineering and Technology, Peshawar, Pakistan, in 2009, the master's degree in satellite navigation from the Polito di Torino, Italy, in 2012, and the Ph.D. degree in electronics and communication engineering from Hanyang University, Seoul, South Korea, in 2015. He was an Assistant Professor with the Department of Electrical Engineering, Gandhara Institute of Science and IT, Peshawar, from August 2015 to September 2016. He has worked as an Assistant Professor with IQRA National University, Peshawar, from October 2016 to July 2017. From July 2017 to December 2020, he was a Postdoctoral Research Fellow with the Communication Theory Laboratory, King Abdullah University of Science and Technology (KAUST). He is currently an Associate Professor with the Department of Electrical Engineering, Northern Border University, Arar 73222, Saudi Arabia. His current research interests include cognitive radio networks, non-conventional wireless communications, aerial networks, dimensionality reduction, and localization.
\end{IEEEbiography}

\begin{IEEEbiography}[{\includegraphics[width=1in, height=1.25in, clip, keepaspectratio]{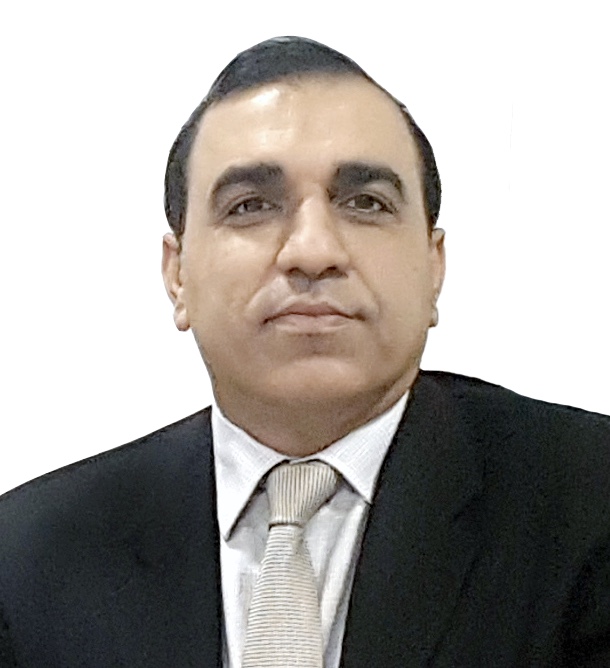}}]
{MOHAMMAD INAYATULLAH BABAR} received his Bachelor of Science Degree in Electrical Engineering from University of Engineering and Technology (UET), Peshawar, Pakistan in 1997.  He received his Masters and Doctorate Degrees in 2001 and 2005 respectively from School of Engineering and Applied Sciences, George Washington University, Washington DC USA.  He is a member of IEEE USA and ACM USA. He also taught a number of Telecommunications Engineering Courses at Graduate Level in School of Engineering, Stratford University, Virginia USA as Adjunct faculty. Currently, he is working as Professor in Department of Electrical Engineering,  supervising postgraduate Scholars in the field of Wireless Communications Network.
\end{IEEEbiography}

\begin{IEEEbiography}[{\includegraphics[width=1in, height=1.25in, clip, keepaspectratio]{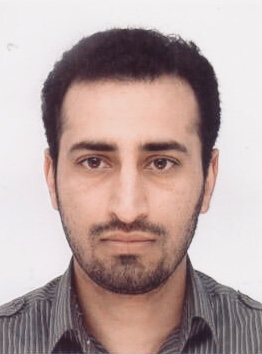}}]
{TARIQULLAH JAN}  did his PhD in the field of Electronic Engineering from the University of Surrey, United Kingdom in 2012. He did his Bachelor in Electrical Engineering from the University of Engineering and Technology Peshawar, Pakistan in 2002. Currently he is serving as Associate Professor at Department of Electrical Engineering, Faculty of Electrical and Computer Systems Engineering, University of Engineering and Technology Peshawar, Pakistan. His Research interest includes Blind signal processing, machine learning, blind reverberation time estimation, speech enhancement, multimodal based approaches for the blind source separation, compressed sensing, and Non-negative matrix/tensor factorization for the blind source separation.
\end{IEEEbiography}

\begin{IEEEbiography}[{\includegraphics[width=1in, height=1.25in, clip, keepaspectratio]{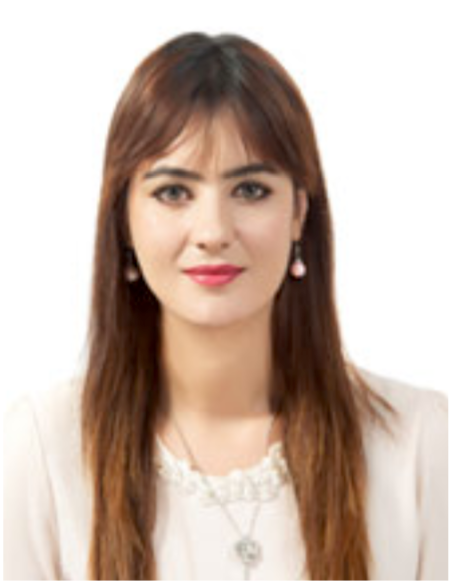}}]
{SADIA DIN}  is currently working as an Assistant Professor in the Department of Information and Communication Engineering, Yeungnam University, South Korea. Dr. Sadia is also working as a senior researcher in University of Milano, Milan, Italy. Previously, she was working as a Post-Doctoral Researcher in Kyungpook National University, South Korea (Mar 2020 ~ Aug 2020). She received her Ph.D. in Data Science and Masters in Computer Science from Kyungpook National University, South Korea, and Abasyn University, Islamabad Pakistan in 2020 and 2015, respectively. During her Ph.D., she was working on various projects including Demosaicking and Denoising using Machine/Deep Learning, Artificial Learning. Furthermore, she extended her research toward Internet of Things, 5G, and Big Data Analytics. At the beginning of her research career, she has published highly more than 60 journals and conferences including IEEE IoT, IEEE TII, IEEE Wireless Communication, IEEE Globecom, IEEE LCN, IEEE Infocom, etc. In addition, she is the recipient of two Korean patents in 2019 and 2020. In 2015, she was visiting researcher at CCMP Lab, Kyungpook National University, South Korea, where she was working on Big Data and Internet of Things. Moreover, she is the recipient of two international awards, i.e., research Internship at CCMP Research Lab, Kyungpook National University, S. Korea (June 2015), and CSE Best Research Award at Kyungpook National University, S. Korea (October 29, 2019). She was also the chair for the IEEE International Conf. on Local Computer Networks (LCN’18). She is serving as a Guest Editor in journal of Wiley, Big Data, and Microprocessor and microsystem. In IEEE LCN 2017 in Singapore, she has chair couple of sessions. Her area of research is Demosaicking and Denoising using Machine/Deep Learning, Artificial Learning, Big Data analytics, 5G, and IoT. Moreover, she got CSE Best Research Award at Kyungpook National University, S. Korea (October 29, 2019.

\end{IEEEbiography}

\end{document}